# Multitasking Deep Learning Model for Detection of Five Stages of Diabetic Retinopathy

Sharmin Majumder, *Graduate Student Member, IEEE*, Nasser Kehtarnavaz, *Fellow, IEEE*

**Abstract** This paper presents a multitask deep learning model to detect all the five stages of diabetic retinopathy (DR) consisting of no DR, mild DR, moderate DR, severe DR, and proliferate DR. This multitask model consists of one classification model and one regression model, each with its own loss function. Noting that a higher severity level normally occurs after a lower severity level, this dependency is taken into consideration by concatenating the classification and regression models. The regression model learns the inter-dependency between the stages and outputs a score corresponding to the severity level of DR generating a higher score for a higher severity level. After training the regression model and the classification model separately, the features extracted by these two models are concatenated and inputted to a multilayer perceptron network to classify the five stages of DR. A modified Squeeze Excitation Densely Connected deep neural network is developed to implement this multitasking approach. The developed multitask model is then used to detect the five stages of DR by examining the two large Kaggle datasets of APTOS and EyePACS. A multitasking transfer learning model based on Xception network is also developed to evaluate the proposed approach by classifying DR into five stages. It is found that the developed model achieves a weighted Kappa score of 0.90 and 0.88 for the APTOS and EyePACS datasets, respectively, higher than any existing methods for detection of the five stages of DR.

## I. INTRODUCTION

INTERNATIONAL Diabetes Federation (IDF) states that there are more than 460 million adults (20-79 years) in the world living with diabetes. The number of adults with diabetes has more than tripled over the past 20 years and 1 in 2 people (about 230 million) with diabetes are undiagnosed [1]. A complication of diabetes is Diabetic Retinopathy (DR), an eye retinal disease that can lead to visual impairments and even blindness. DR is caused by microvascular complications of diabetes appearing as morphological changes in the eye fundus. Diagnosis and treatment of DR at its initial stages reduces the risk of vision loss to a great extent. With the development of color fundus photography, DR can be detected non-invasively at its early stages.

Along with the detection of DR, the severity level of DR also needs to be determined for treatment purposes. There are two major types or classes of DR: non-proliferative DR (NPDR) and proliferative DR (PDR) [2]. NPDR is further categorized into the following three stages: (i) mild NPDR, which is the earliest stage of DR, (ii) moderate NPDR, and (iii) severe NPDR. PDR denotes the advanced stage of DR. The severity level of DR is thus generally graded as these five stages: no DR, mild DR, moderate DR, severe DR, and proliferate DR.

Lesions in fundus images that appear as small circular red dots at the end of blood vessels indicate the earliest sign of DR. Microaneurysms, Hemorrhages and/or Exudates are signs of modeate DR. In the PDR stage, new blood vessesls are formed along with the above abnormalities [3]. Figure 1 shows sample color fundus images of normal retina and different severity levels of DR. A major issue with DR detection involves the difficlty of identifying symptoms at its early stages due to visual similarities between no DR, mild DR, and sometimes moderate DR. If DR proceeds to the advanced stage, vision loss can occur.

Many computer-based methods have been developed in the litreature for the detection of DR. In these previous methods, to mimick human experts, much attention has been paid to the automatic detection of lesions for DR screening and grading. A representative set of methods already developed are stated here. Detection and segmentation of blood vessels in retinal images was discussed in [4], [5]. In case of Microaneurysm detection, automated image processing approaches were developed in [6]–[9]. Several methods for detection of Exudates in color fundus images were introduced in [10]–[13]. In [14], the detection of neovascularisation and lesions was performed for the Messidor dataset.

Lately, due to the success of deep learning models in many image processing tasks, researchers have utilized them for DR detection. In [15], a region-based fully convolutional network (R-FCN) for lesion detection and DR grading into four stages was developed. In [16], an instance learning was used to detect lesions in fundus images for the Messidor dataset.

There have also been some works reported on detecting DR stages, that is conducting detection as well as

The authors are with the Department of Electrical and Computer Engineering, University of Texas at Dallas, Richardson, TX 75080, USA (email: sharmin.majumder@utdallas.edu, kehtar@utdallas.edu .

classification of DR at the same time. These types of image classification tasks can be grouped into conventional image processing techniques where handcrafted features were considered [17]–[24], and more recent deep learning techniques [25]–[27]. In [21], an algorithm based on random forest was applied to handcrafted features to detect the presence of DR and assess its risk. In [22], a DR classification was performed by using BossaNova and Fisher Vector midlevel features which extended the classical Bags of Visual Words features. In [23], a two-step method based on handcrafted features was covered: one step for detecting the presence of DR and one step for detecting its severity level. A bag of features approach was developed for detection of DR stages by using the histogram of orientated gradients in [24]. In [28], both binary and multiclass classification of DR was achieved by using Haralick and multiresolution features.

Deep learning techniques, especially convolutional neural networks (CNNs), have generated much success in image classification due to their end-to-end learning capabilities or not requiring to devise handcrafted features [29]–[31]. In [27], three CNN models were utilized for the binary classification of DR, that is DR/no DR. Deep learning-based classification approaches were also discussed in [26], [32]–[35]. [26], [34], [35] focused on binary classification of DR as referable and non-referable. In [35], the EfficientNet-B5 model was used for this classification task. In [36], binocular fundus images from both eyes (left and right eye) were taken as the inputs to a transfer learning-based CNN model. In [37], a DCNN (Deep Convolution Neural Network) for detecting two stages of DR (normal and NPDR) was discussed. In [38], the right and left eye images were treated separately by applying CNN models for binary classification of DR. In [25], a CNN based smartphone app was developed for binary classification of DR in real-time.

Classification of the severity stages of DR were presented in [39]–[41]. In [42], a deep neural network for four-degree classification of DR was covered. A hyperparameter tuning was done in the Inception-v4 model to obtain four classes of DR in [39]. A CNN model was developed to classify the five stages of DR in [40]. In [43], three deep learning models (Feed Forward Neural Network (FNN), Deep Neural Network (DNN), and Convolutional Neural Network (CNN) were applied to the EyePACS dataset for DR classification whereas the performance of the EyePACS dataset was examined for different CNN models in [43]. In [44], several deep learning models (AleXNet, VggNet, GoogleNet, ResNet) were compared for DR classification using the Kaggle EyePACS dataset with VggNet achieving the best accuracy. We developed a transfer learning-based smartphone app using a pretrained Xception model for classification of the five stages of DR in real time [41].

A few recent papers have utilized ensembles of two or more deep learning models for DR classification. In [45], the integration of deep learning models was used to detect no DR, referable DR (rDR), vision threatening DR, and macular edema. In [46], an ensemble of five pretrained CNN models consisting of Resnet50, Inceptionv3, Xception, Dense121, and Dense169, were used for DR classification into five stages. All these papers considered ensemble of two or more classification models but did not employ any regression task. Moreover, the above mentioned or existing five-stage DR classification papers have reported not high accuracy when considering all the stages. A few papers in the literature also proposed multitasking network for fundus image analysis mainly focusing on lesion segmentation task [47]–[49]. In [47], a weakly supervised multitask architecture is proposed for retinal lesions segmentation whereas simultaneous segmentation of bright and red lesions in fundus images is performed using a multitasking architecture in [48]. In [49], a region-specific multitask recognition model was proposed to classify 36 different retinal disease without examining the classification of different DR stages.

In this paper, a multitasking deep learning architecture is proposed for classifying fundus images into the five stages of DR (no DR, mild DR, moderate DR, severe DR, and proliferate DR). This multitasking model learns the inter-dependency among the different stages along with the distinctions between the stages performing regression and classification tasks, respectively. A MultiLayer Perceptron (MLP) model is used at the end to classify the five stages of DR based on the features extracted from the two networks (classification and regression). A densely connected network modified with squeeze excitation (SE) layers is developed to implement the proposed multitasking method due to the capability of SE layers to learn channel interdependencies at almost no computational cost. Since deep neural networks require a large amount of data to train from the scratch, we also evaluated our proposed multitasking approach by an Xception transfer learning model. The two large public domain Kaggle datasets, namely APTOS [50] and EyePACS [51], are examined in this work to evaluate the performance of the proposed approach.

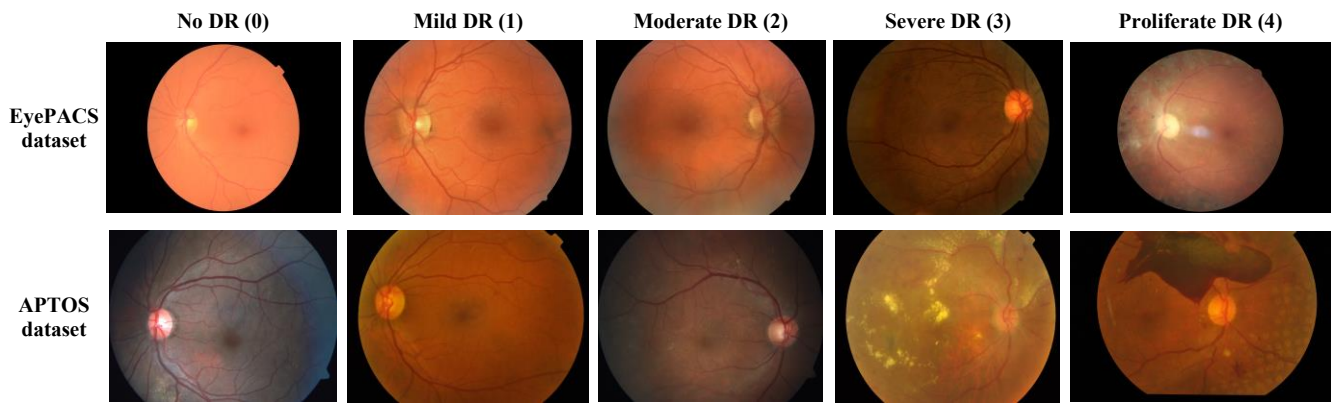

Fig. 1. Sample fundus images of the five stages of DR

The rest of the paper is organized as follows: a description of the datasets considered together with a description on the proposed multitasking approach are provided in Section II. The developed multitasking squeeze excitation densely connected model and multitasking Xception transfer learning model are then presented in Section III. Section IV describes the experimentations carried out together with their results followed by the conclusion in Section V.

## II. MATERIALS AND METHODS

### A. Datasets

Choice of dataset is important as it needs to contain a rich collection of images. In this work, two publicly available Kaggle datasets that incorporate a large number of images of all the five stages of DR are considered. These datasets are EyePACS (Eye Picture Archive Communication System) [51] and APTOS 2019 (Asia Pacific Tele-Ophthalmology Society) Blindness Detection Dataset [50].

*APTOS 2019 Blindness Detection dataset:* This dataset includes fundus images for the five stages of DR labeled by the severity level 0 to 4, where label 0 indicates no DR, label 1 mild DR, label 2 moderate DR, label 3 severe DR, and label 4 proliferate DR. It contains a total of 3,662 retinal images where 1,805 images belong to no DR, 370 images to mild DR, 999 to moderate DR, 193 images to severe DR, and 295 images to proliferate DR. The resolution of the images is 3216×2136.

*EyePACS dataset:* Similar to the APTOS dataset, the EyePACS dataset also contains fundus images belonging to the five stages of DR. This dataset contains 35,126 retina images of size 3888×2951 for both the left and right eyes, with 25,810 images labeled as 0 DR (no DR), 2,443 mild DR, 5,292 moderate DR, 873 severe DR, and 708 proliferate DR images. Here, 10,000 images are randomly selected from the no DR stage and our model is trained on a total 19,316 images (10,000 no DR, 2,443 mild DR, 5,292 moderate DR, 873 severe DR, and 708 proliferate DR).

It is to be noted that the above datasets are highly imbalanced thus introducing bias. To have a balanced dataset, a class weighting method is applied to weigh classes inversely proportional to their frequency according to Eq. (1)

$$w_j = \frac{n}{k n_j} \quad (1)$$

where $w_j$ is the weight of class $j$, $n$ is the total number of samples, $n_j$ is the number of samples in class $j$, and $k$ is the total number of classes.

### B. Data Preprocessing and Augmentation

The datasets considered in this paper contain images of high resolution. The images are resized to 299×299 to feed into the networks. After resizing, input image intensity values are normalized between 0 and 1. Deep learning is data hungry. The amount of data in the above two datasets are not sufficient for training a deep neural network from scratch. Therefore, data augmentation techniques such as rotation, horizontal flip, width shift, height shift, zooming, and shearing are applied to the original data.

### C. Proposed Multitasking Method

Classification task usually works based on the difference or distinction between the classes. This type of task uses one loss function. Sometimes, more than one loss function improves the classification performance [52]. Diseases like diabatic retinopathy progress to higher severity stages from lower severity stages. For example, severe DR stage comes after moderate DR stage, moderate DR stage comes after mild DR stage, etc., therefore leads to a dependency among different stages. This dependency characteristics between the stages can be learnt by a regression model which contributes to further improving the classification task. Keeping this in mind, we propose a multitasking model consisting of a classification and a regression model to classify or detect the five stages of DR, see Fig. 2.

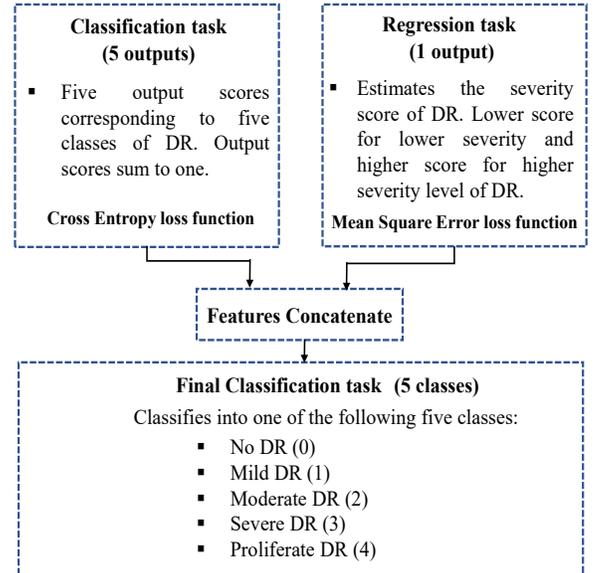

Fig. 2. Developed multitasking approach.

The classification model learns the distinguishing characteristics between the five stages whereas the regression model learns the inter-dependency characteristics among the five stages. Two loss functions are considered, *cross entropy* loss function as given in Eq. (2) is used for classification task and *mean square error* loss function as defined by Eq. (3) is used for regression task. The regression model and the classification model are trained separately on the fundus images using a linear activation function and a softmax activation function (Eq. (5)) in the last layer of the regression model and the classification model, respectively.

The classification model outputs five probability scores (sum to one) corresponding to the five classes or stages of DR. The regression model scores one output corresponding to the severity level of DR. The regression model is trained with output labels 0 for no DR, 0.2 for mild DR, 0.4 for moderate DR, 0.6 for severe DR, and 0.8 for proliferate DR. Features learnt by the classification model and the regression model are concatenated and fed into an MLP classifier for the final classification of the five stages of DR. The proposed approach is presented as an algorithm in Algorithm 1.

$$Cross\ entropy = -\sum_{i}^{M} y_i \log(y_i) \quad (2)$$

$$MSE = \frac{1}{M}\sum_{i=1}^{M}(y_i - y_i)^2 \quad (3)$$

where, $y_i$ is the predicted value and $y$ is the true value. $M$ is the number of classes.

$$Soft\max(i) = \frac{e^{\delta(i)}}{\sum_{j=1}^{K} e^{\delta(j)}},\ j = 1,..i,..K \quad (4)$$

where $K$ denotes the total number of classes, and $\delta$ denotes the output of the last fully connected layer. The output probabilities of each class lie between 0 and 1 with all the values adding up to 1.

---

**Algorithm 1:** Algorithm of multitasking approach (lr=learning rate, $\beta_1,\beta_2$=exponential decay rate in Adam optimization for the first moment and second moment estimates, respectively)

**Require:** Fundus Images and Labels $(X,Y)$, where $Y = \{y\ /\ y \in \{0,1,2,3,4\}\}$ [0: No DR, 1: Mild DR, 2: Moderate DR, 3: Severe DR, 4: Proliferate DR]

**Input:** fundus images $x \in X$

**Output:** Trained model predicts probability scores corresponding to $\forall y$ for an input $x$

Perform Preprocessing:
- Resize the image to $299 \times 299 \times 3$
- Perform Data Augmentation techniques: rotation, horizontal flip, width shift, height shift, zooming, and shearing.

Design a Classification Model and a Regression Model $H = \{Classification\ Model, Regression\ Model\}$

**for** $\forall h \in H$ **do**
    lr=0.001, momentum=0.7
    **for** epochs=1 to 250 **do**
        **for** each minibatch $(X_{mini}, Y_{mini}) \in (X,Y)$ **do**
            **if** $h = Classification\ Model$ **then**
                Update the parameters of the Classification Model using Stochastic Gradient Decent optimization (SGD).
                **if** epochs>150 **then**
                      lr=0.0001
                **end**
                **if** epochs>200 **then**
                      lr=0.00001, momentum=0.5
                **end**
            **end**
            **if** $h = Regression\ Model$ **then**
                **if** epochs<50 **then**
                      Update the parameters of the Regression Model using Adaptive Moment Estimation (Adam).
                **end**
            **end**
        **end**
    **end**
**end**

Concatenate the features extracted from $\forall h \in H$ and fed to a MLP classifier to generate the Multitasking Model.

**for** Multitasking Model **do**
    lr=0.001, $\beta_1$=0.9, beta_$\beta_2$=0.999
    **for** epochs=1 to 50 **do**
        **for** each minibatch $(X_{mini}, Y_{mini}) \in (X,Y)$ **do**
            Update the parameters of the Multitasking Model using Adaptive Moment Estimation.
            **if** the validation error is not improving for four epochs **then**
                lr = lr × 0.01
            **end**
        **end**
    **end**
**end**

**for** $x \in X_{test}$ **do**
    Trained Multitasking Model predicts probability scores for $\forall y$
**end**

## III. Implementation of multitasking Deep Learning Models

### A. *Squeeze Excitation Densely Connected Multitasking Network (MSEDenseNet)*

A modified densely connected network (DenseNet) is developed here to implement the multitasking approach. A basic DenseNet with compression is combined with a squeeze-excitation (SE) network. SE network introduces a building block that improves channel interdependencies to improve the performance of the model. The Multitasking Squeeze Excitation Densely Connected Network (MSEDenseNet) consists of a SEDenseNet classification model, a SEDenseNet regression model, and a MLP classifier. The architecture of the developed MSEDenseNet is shown in Fig. 3.

*1) Model Architecture*

As shown in Fig. 3, SEDenseNet consists of five dense blocks and four transition blocks each in between two dense blocks. In each dense block, a SE-dense module has been repeated for 16 times. A SE-dense module consists of a batch normalization layer, ReLU activation, a 3x3 convolution layer, and a SE block. A SE block comprises a squeeze layer which is a global average pooling layer, and an excitation layer with two 1x1 convolution layers. The first convolution layer is followed by a ReLU activation and the second convolution layer is followed by a sigmoid activation. In SE block, each channel is squeezed to a single numeric value using average pooling. The ratio to reduce the channel complexity is set to 16. Finally, each channel of the input to the SE block is scaled by the respective weight obtained from the SE block. Down-sampling is achieved by the transition layer between two dense blocks. A transition layer is made of batch normalization, ReLU, 1x1 convolution, and average pooling. A SE block is also added to the transition layer. The last fully connected layer of the original network is replaced with 2x2 convolution layer to reduce the number of parameters.

The depth and growth rates of the developed SEDenseNet network is set to 164 and 18, respectively. Therefore, the number of dense modules in one dense block is 16. The number of filters for the first convolution layer is $2 \times$ growth rate whereas for the convolution layers in the dense block and in the transition block are $2 \times$ growth rate $\times$ compression ratio. The compression ratio for the network is set to 0.5.

In the SEDenseNet multitasking model, a SEDenseNet classification model, and a SEDenseNet regression model are combined to enrich the learned features. Fig. 3 illustrates the concatenation of the regression model and the classification model. Outputs from the last average pooling layer of the pretrained classification model and the regression model are fused together to feed into the MLP. The MLP comprises a batch normalization layer, a fully connected layer of 512 units with ReLU activation, and another fully connected layer with softmax activation function as the output layer.

*2) Model Training*

The developed SEDenseNet classification model is trained with 250 epochs using Stochastic gradient descent (SGD) optimization algorithm and Categorical Cross-entropy (CCE) loss function. For the first 150 epoch, the learning rate, batch size, and momentum are set to 0.001, 2, and 0.7, respectively. For the next 50 epochs, the learning rate is reduced to 0.0001. For the last 50 epochs, the learning rate and momentum are changed to 0.00001 and 0.5, respectively. The Validation accuracy is checked in every epoch and the model with the highest validation accuracy is saved using the model checkpoint feature of the Keras callback. The output layer of the classification model is a convolution layer with softmax activation function to generate five probability scores corresponding to the five classes of DR.

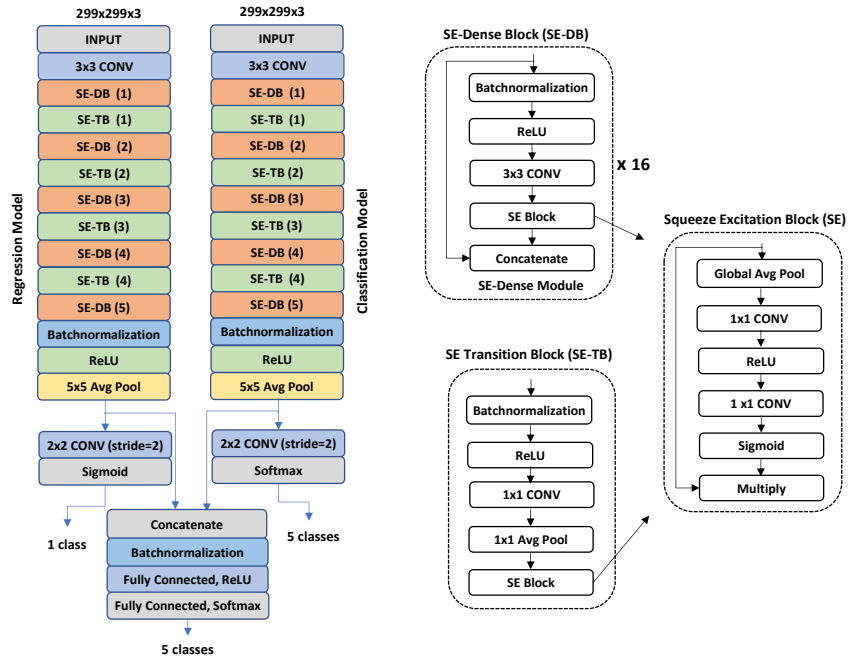

Fig. 3. Architecture of the developed Multitasking Squeeze Excitation Densely Connected Deep Neural Network (MSEDenseNet). (CONV: Convolution)

The developed SEDenseNet regression model is trained with 50 epochs. Adaptive moment estimate (Adam) optimization algorithm with a learning rate of 0.001 and mean square error (MSE) loss function are used to train the model. The mini batch size is kept 2. The output layer of the regression model is a convolution layer with linear activation function to score one output corresponding to the severity level of DR.

After concatenation of the features from the previously trained classification and regression model, a batch normalization layer with momentum 0.9 is added for the purpose of normalizing data across a batch. A batch normalization layer is usually used to speed up the training and to reduce the sensitivity to initialization. The CCE loss function and the Adam optimization algorithm with a learning rate of 0.001 are considered during training for 50 epochs. The learning rate is reduced by a factor of 0.1 if the validation loss is not reduced for four consecutive epochs. The model which has the highest validation accuracy is then saved.

The so-called 'He normal' initialization and 'l2 kernel' regularization are considered for the units of the convolution and fully connected layers of MSEDenseNet. The regularization protects the model against overfitting.

B. *Multitasking Xception Transfer Learning Model (MXception)*

Due to the scarcity of sufficient training data to train a deep neural network from scratch, a widely used model with transfer learning is also considered. A pretrained Xception ImageNet model is fine-tuned to implement the multitasking approach to classify the five stages of DR.

*1) Model Architecture*

The architecture of the Xception model [53] is based on depthwise separable convolution layers and consists of three major sections: entry flow, middle flow, and exit flow. Fig. 4 shows the architecture of the Xception model. Image data first goes through the entry flow, then through the middle flow which is repeated eight times, and finally through the exit flow. Note that all the convolution and separable convolution layers are followed by batch normalization. This model is composed of 36 convolutional layers forming the feature extraction base of the network. The Xception model was previously trained with 299×299 ImageNet images for 1000 classes with the top-1 accuracy of 79%.

*2) Model Fine Tuning*

A pretrained Xception ImageNet model is fine-tuned as a regression model with one class in the output. The architecture of the regression Xception model is shown in Fig. 4. The last fully connected layer of the Xception model is chopped and then an average pooling layer is added. A dense layer consisting of one neuron is also added as the output layer with linear activation function. The adaptive moment estimate (Adam) optimization algorithm with a learning rate of 0.001 and MSE loss function are used to train the model for 25 epochs. During training, the image dataset is split into mini batches of size 16.

Another pretrained Xception model is fine-tuned with retinal images to classify the five stages of DR. The last fully connected layer is replaced with an average pooling layer, and a dense layer with softmax activation function. A dropout layer is also added before the output layer with 0.8 keep probability to regularize the model. The CCE loss function and Adam optimization with a learning rate of 0.001 (0.9 exponential decay rate for the first-moment estimates, β1 and 0.999 exponential decay rate for the second-moment estimates, β2) are used for training. Model with the highest validation accuracy is saved. The model is fine tuned for 25 epochs with minibatch size of 16. If the validation loss is not reduced for four consecutive epochs, the learning rate is reduced by a factor of 0.1.

Features generated from the last average pooling layer of the fine-tuned Xception classification model and regression model are concatenated and inputted to an MLP classifier.

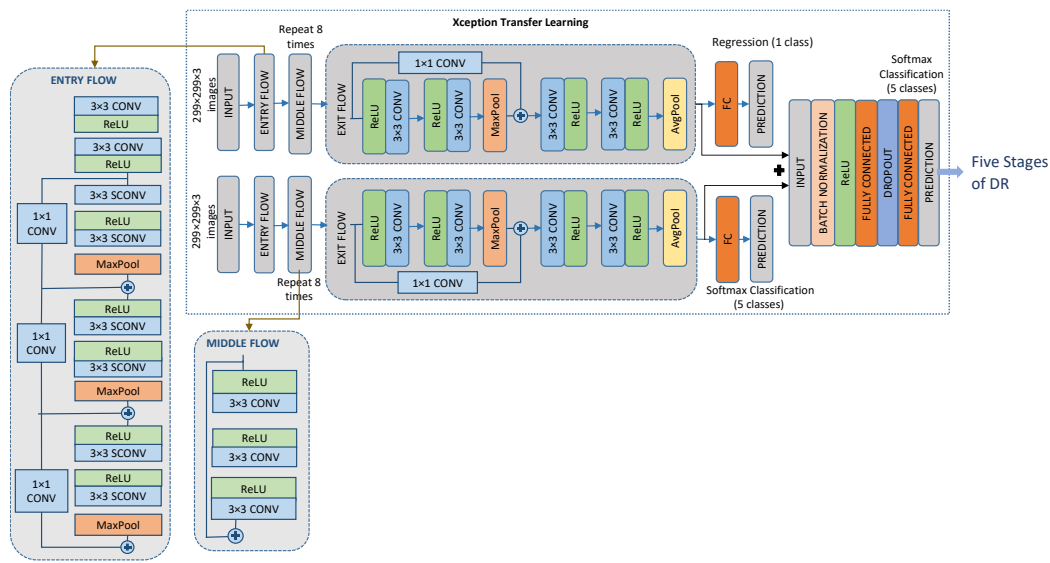

Fig. 4. Architecture of the developed multitasking Xception transfer learning model (MXception) - CONV: Convolution, FC: Fully Connected Layer, SCONV: Separable Convolution.

Similar to the SEDenseNet, MLP in the Multitasking Xception network also contains two fully connected layers with softmax activation function at the last layer to generate five score for the five classes or stages of DR. The training parameters to train the MLP classifier in the Xception multitasking network are similar to the parameters used in the MLP classifier of MSEDenseNet. The model which has the highest validation accuracy is then saved.

## IV. EXPERIMENTATIONS AND RESULTS

### A. Implementation Framework

The experimentations reported here were carried out on a computer equipped with an NVIDIA Quadro P5000 GPU. The computer had an Intel® Core™ i9 processor with twenty 3.3GHZ cores and 32GBs of RAM. The software packages used for implementation of the models included Python 3.7 together with the deep learning libraries of Keras with Tensorflow, H5PY, OpenCV, and Scikit-Learn.

### B. Performance measures

Performance was assessed based on the five widely used performance measures of *Precision, Recall, F1 Score, Accuracy,* and *Weighted Kappa Score (WKS)* as stated in Eqs. (5), (6), (7), (8), and (9), respectively. *Precision*, *Recall*, and *Accuracy* are computed based on True Positive (*TP*), True Negative (*TN*), False Positive (*FP*), and False Negative (*FN*). *TP* indicates correctly classified positive classes, *FP* indicates negative classes misclassified as positive, *FN* indicates positive classes misclassified as negative, and finally true negatives (*TN*) indicates correctly classified negative classes. *Precision* and *Recall* were first computed for the five classes separately and then a macro average was taken for the multiclass classification. In case of *Kappa Score,* Quadratic weight was considered.

$$Precsion = \frac{TP}{TP+FP} \quad (5)$$

$$Recall = \frac{TP}{TP+FN} \quad (6)$$

$$F1\ Score = 2*\frac{Precision*Recall}{Precision+Recall} \quad (7)$$

$$Accuracy = \frac{TP}{TP+FN+TN+FP} \quad (8)$$

$$WKS = 1 - \frac{\sum_{i=1}^{N}\sum_{j}^{N}W_{ij}O_{ij}}{\sum_{i=1}^{N}\sum_{j}^{N}W_{ij}E_{ij}} \quad (9)$$

where $i$ and $j$ denote the indices associated with the true class and the classified class, respectively, $O$ is actual observation counts, $E$ is expected counts, $N$ is the total number of classes, and $W_{ij}$ is given by Eq. (10):

$$W_{ij} = \frac{(i-j)^2}{(N-1)^2} \quad (10)$$

### C. Classification Outcome

*Results for APTOS dataset:* For APTOS dataset, three different experiment sets were considered by randomly selecting 90% of the data for training and the remaining 10% for validation. Confusion matrices of the validation datasets for experiment 1 using the SEDenseNet Classification and SEDenseNet multitasking model are shown in TABLE I, and TABLE II, respectively. The performance measures of *Precision, Recall*, *F1 Score*, *Accuracy*, and *Quadratic weighted Kappa Score* for experiment 1 are shown in TABLE III. As can be seen from this table, the multitasking model improved the classification performance by nearly 4%.

TABLE I
CONFUSION MATRIX ON THE APTOS DATASET FOR SEDENSENET CLASSIFICATION MODEL: EXP 1

| | | Classified | | | | |
|---|---|---|---|---|---|---|
| | | Not DR | Mild DR | Moderate DR | Severe DR | Proliferate DR |
| Actual | Not DR | 183 | 3 | 2 | 0 | 0 |
| | Mild DR | 4 | 15 | 12 | 0 | 2 |
| | Moderate DR | 3 | 5 | 82 | 2 | 8 |
| | Severe DR | 0 | 0 | 10 | 4 | 1 |
| | Proliferate DR | 0 | 2 | 15 | 1 | 13 |

TABLE II
CONFUSION MATRIX ON THE APTOS DATASET FOR SEDENSENET MULTITASK MODEL: EXP 1

| | | Classified | | | | |
|---|---|---|---|---|---|---|
| | | Not DR | Mild DR | Moderate DR | Severe DR | Proliferate DR |
| Actual | Not DR | 183 | 4 | 1 | 0 | 0 |
| | Mild DR | 3 | 24 | 5 | 0 | 1 |
| | Moderate DR | 2 | 7 | 83 | 4 | 4 |
| | Severe DR | 0 | 0 | 7 | 8 | 0 |
| | Proliferate DR | 0 | 2 | 13 | 1 | 15 |

TABLE III
PERFORMANCE MEASURES OF THE DEVELOPED SEDENSENET MODELS FOR EXP 1: APTOS DATASET

| Models | Precision | Recall | F1-score | Accuracy | WKS |
|---|---|---|---|---|---|
| SEDenseNet Classification | 0.67 | 0.59 | 0.61 | 0.81 | 0.84 |
| SEDenseNet Multitask | 0.75 | 0.70 | 0.72 | 0.85 | 0.88 |

The confusion matrices for the experiment 2 using the two models are shown in TABLE IV and TABLE V, respectively whereas the performance measures are shown in TABLE VI. This table also shows the improvement in the classification performance for the multitasking model.

TABLE IV
CONFUSION MATRIX ON THE APTOS DATASET FOR SEDENSENET CLASSIFICATION MODEL: EXP 2

| | | Classified | | | | |
|---|---|---|---|---|---|---|
| | | Not DR | Mild DR | Moderate DR | Severe DR | Proliferate DR |
| Actual | Not DR | 176 | 5 | 0 | 0 | 0 |
| | Mild DR | 4 | 11 | 21 | 0 | 1 |
| | Moderate DR | 2 | 3 | 94 | 1 | 4 |
| | Severe DR | 0 | 0 | 11 | 2 | 5 |
| | Proliferate DR | 1 | 1 | 9 | 0 | 16 |

TABLE V
CONFUSION MATRIX ON THE APTOS DATASET FOR SEDENSENET MULTITASK MODEL: EXP 2

| | | Classified | | | | |
|---|---|---|---|---|---|---|
| | | Not DR | Mild DR | Moderate DR | Severe DR | Proliferate DR |
| Actual | Not DR | 174 | 5 | 2 | 0 | 0 |
| | Mild DR | 3 | 25 | 9 | 0 | 0 |
| | Moderate DR | 0 | 11 | 88 | 4 | 1 |
| | Severe DR | 0 | 1 | 6 | 6 | 5 |
| | Proliferate DR | 1 | 3 | 8 | 0 | 15 |

TABLE VI
PERFORMANCE MEASURES OF THE DEVELOPED SEDENSENET MODELS FOR EXP 2: APTOS DATASET

| Models | Precision | Recall | F1-score | Accuracy | WKS |
|---|---|---|---|---|---|
| SEDenseNet Classification | 0.69 | 0.58 | 0.63 | 0.81 | 0.87 |
| SEDenseNet Multitask | 0.725 | 0.67 | 0.70 | 0.84 | 0.88 |

The confusion matrices for experiment 3 in case of the classification and multitasking SEDenseNet are presented in TABLE VII and TABLE VIII, respectively. The performance measures for this experiment are shown in TABLE IX indicating improved classification performance in the multitasking case.

TABLE VII
CONFUSION MATRIX ON THE APTOS DATASET FOR SEDENSENET CLASSIFICATION MODEL: EXP 3

| | | Classified | | | | |
|---|---|---|---|---|---|---|
| | | NDR | MDR | MoDR | SDR | PDR |
| Actual | Not DR | 170 | 4 | 3 | 0 | 1 |
| | Mild DR | 4 | 20 | 15 | 0 | 3 |
| | Moderate DR | 2 | 4 | 91 | 2 | 1 |
| | Severe DR | 0 | 0 | 10 | 8 | 2 |
| | Proliferate DR | 0 | 2 | 11 | 2 | 12 |

TABLE VIII
CONFUSION MATRIX ON THE APTOS DATASET FOR SEDENSENET MULTITASK MODEL: EXP 3

| | | Classified | | | | |
|---|---|---|---|---|---|---|
| | | NDR | MDR | MoDR | SDR | PDR |
| Actual | Not DR | 171 | 4 | 3 | 0 | 0 |
| | Mild DR | 2 | 31 | 7 | 0 | 2 |
| | Moderate DR | 3 | 6 | 88 | 2 | 1 |
| | Severe DR | 0 | 1 | 9 | 8 | 2 |
| | Proliferate DR | 0 | 2 | 10 | 0 | 15 |

TABLE IX
PERFORMANCE MEASURES OF THE DEVELOPED SEDENSENET MODELS FOR EXP 3: APTOS DATASET

| Models | Precision | Recall | F1-score | Accuracy | WKS |
|---|---|---|---|---|---|
| SEDenseNet Classification | 0.79 | 0.71 | 0.74 | 0.85 | 0.87 |
| SEDenseNet Multitask | 03 | 0.83 | 0.87 | 03 | 0.83 |

In addition, the Xception transfer learning was examined to show the effectiveness of the multitasking method. 90% of the data were randomly selected for fine tuning the model and the remaining 10% were used for validation. The performance of Xception Classification and Xception Multitask models along with the performance of the SEDenseNet Classification and SEDenseNet Multitask model averaged over the three experiments are shown in TABLE X. As can be seen from this table, the multitasking model improved the classification performance by nearly 3% when using the SEDenseNet and the Xception models.

TABLE X
PERFORMANCE MEASURES OF THE DEVELOPED MODELS: APTOS DATASET

| Models | Precision | Recall | F1-score | Accuracy | WKS |
|---|---|---|---|---|---|
| SEDenseNet Classification | 0.70 | 0.60 | 0.64 | 0.81 | 0.85 |
| SEDenseNet Multitask | 0.76 | 0.69 | 0.72 | 0.85 | 0.88 |
| Xception Classification | 0.74 | 0.68 | 0.70 | 0.83 | 0.87 |
| Xception Multitask | 0.77 | 0.70 | 0.73 | 0.86 | 0.90 |

*Results for EyePACS dataset:* The multitasking approach was also applied to the EyePACS dataset. For this dataset, only the Xception multitasking transfer learning model was considered. The model was fine tuned using a total of 19,316 images from the EyePACS dataset. 80% of the images were randomly selected for training and the remaining 20% were used for validation. The performance measures of *Recall*, *Precision*, *F1 Score*, *Accuracy*, and *Quadratic weighted Kappa Score* for the Xception multitask model are presented in TABLE XI.

TABLE XI
PERFORMANCE MEASURES OF XCEPTION MULTITASK MODEL: EYEPACS DATASET

| Models | Recall | Precision | F1-score | Accuracy | WKS |
|---|---|---|---|---|---|
| Xception Multitask | 0.64 | 0.69 | 0.66 | 0.82 | 0.88 |

D. *Comparative Study*

The results of the developed MSEDenseNet model for the APTOS dataset was compared with four recent works where the same APTOS dataset was used. The comparison of the performance measures is shown in TABLE XII. As can be seen from this table, the developed multitasking model generated the highest performance measures for the detection of the five stages of DR. TABLE XIII shows the comparison of the performance measures of the multitasking model for the EyePACS dataset with two recent works by Pratt et al. [40] and Qummar et al. [46] where the EyePACS dataset was used. This table also shows the highest performance measures were obtained by the developed multitasking model.

TABLE XII
COMPARISON WITH RECENT WORKS: APTOS DATASET

| Model | Year | Specificity | Accuracy | WKS, KS | No. of Class |
|---|---|---|---|---|---|
| Narayanan et al. [40] | 2020 | - | 96.3 | - | 4 |
| Bodapati et al. [54] | 2020 | - | 80.96 | KS: 0.71 | 5 |
| Kassani et al. [55] | 2019 | 87 | 83.09 | - | 5 |
| Shaban et al. [56] | 2020 | 94-95 | 88-89 | WKS: 0.91-0.92 | 3 |
| **MSEDenseNet** | 2021 | - | 85 | WKS: 0.88 | 5 |
| **MXception** | 2021 | - | 86 | WKS: 0.90 | 5 |

TABLE XIII
COMPARISON WITH RECENT WORKS: EYEPACS DATASET

| Model | Year | Recall | Precision | Specificity | F1 Score | Accuracy | Class No. |
|---|---|---|---|---|---|---|---|
| Pratt et al. [40] | 2016 | 30 | 51.06 | 95 | 41.6 | 75 | 5 |
| Qummar et al. [46] | 2019 | 51.5 | 63.8 | 86.7 | 53.7 | 80.8 | 5 |
| **MXception Model** | 2021 | 64 | 69 | - | 66 | 82 | 5 |

## V. CONCLUSION

Diabetes is a fast-growing disease and there is a 30% chance for a person having diabetes to get Diabetic Retinopathy (DR). DR has different stages from mild to severe and then PDR (Proliferative Diabetic Retinopathy). Computer-based techniques for automatic detection of DR and its different stages have been developed in the literature. This paper has presented a new approach to classify all the five stages of DR from fundus images by using a multitasking deep neural network architecture. The largest publicly available datasets of fundus images (EyePACS and APTOS datasets) were used to train and evaluate the developed ensemble model. The results show that the multitasking model generated the highest performance measures compared to the existing five-stage DR classification methods.

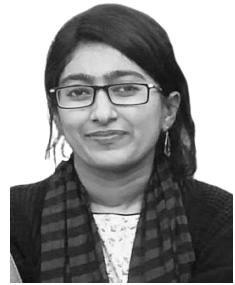

**Sharmin Majumder** (S'17) received the BS and MS degrees in Electrical and Electronic Engineering from the Chittagong University of Engineering and Technology (CUET), Chittagong, Bangladesh, in 2013 and 2019, respectively. She joined the Department of EEE at CUET as a full-time faculty member. She is currently pursuing the PhD degree in Electrical Engineering at the University of Texas at Dallas, Richardson, TX. Her research interests include signal and image processing, ultrasound elastography imaging, computer vision, pattern recognition, and machine learning.

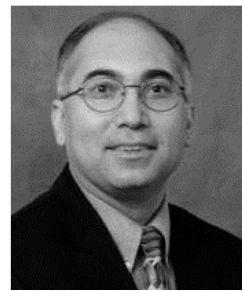

**Nasser Kehtarnavaz** (S'82–M'86–SM'92–F'12) is an Erik Jonsson Distinguished Professor with the Department of Electrical and Computer Engineering and the Director of the Embedded Machine Learning Laboratory at the University of Texas at Dallas, Richardson, TX. His research interests include signal and image processing, machine learning and deep learning, and real-time implementation on embedded processors. He has authored or co-authored 10 books and more than 400 journal papers, conference papers, patents, manuals, and editorials in these areas. He is a Fellow of SPIE, a licensed Professional Engineer, and Editor-in-Chief of *Journal of Real-Time Image Processing*.